\documentclass[aps,pre,preprint,groupedaddress,showpacs,showkeys]{revtex4-1}
\usepackage{graphicx}
\usepackage{amssymb, amsmath}
\begin{document}

\title{1D analysis of 2D isotropic random walks}
\author{Claus Metzner}
\email[]{claus.metzner@gmx.net}
\affiliation{Biophysics Group, University of Erlangen}

\date{\today}

\begin{abstract}
Many stochastic systems in physics and biology are investigated by recording the two-dimensional (2D) positions of a moving test particle in regular time intervals. The resulting sample trajectories are then used to induce the properties of the underlying stochastic process. Often, it can be assumed {\em a priori} that the underlying discrete-time random walk model is independent from absolute position (homogeneity), direction (isotropy) and time (stationarity), as well as ergodic.\\In this article we first review some common statistical methods for analyzing 2D trajectories, based on quantities with built-in rotational invariance. We then discuss an alternative approach in which the two-dimensional trajectories are reduced to one dimension by projection onto an arbitrary axis and rotational averaging. Each step of the resulting 1D trajectory is further factorized into sign and magnitude. The statistical properties of the signs and magnitudes are mathematically related to those of the step lengths and turning angles of the original 2D trajectories, demonstrating that no essential information is lost by this data reduction. The resulting binary sequence of signs lends itself for a pattern counting analysis, revealing temporal properties of the random process that are not easily deduced from conventional measures such as the velocity autocorrelation function.\\In order to highlight this simplified 1D description, we apply it to a 2D random walk with restricted turning angles (RTA model), defined by a finite-variance distribution $p(L)$ of step length and a narrow turning angle distribution $p(\phi)$, assuming that the lengths and directions of the steps are independent.
\end{abstract}
\maketitle

\section*{Quantifying 2D trajectories}

We consider a measured trajectory (Fig.1) that consists of $N+1$ discrete two-dimensional points $\vec{R}_t=(x_t,y_t)$ with $t=0\ldots N$, sampled in equal time intervals $\delta t_{rec}$. In the following, it is implicitly assumed that all absolute times $t$ and lag times $\Delta t$ are in units of this sampling interval. In a spatially homogeneous system, the absolute positions $\vec{R}_t$ are of no importance by themselves. All relevant information about the walk is contained in the N steps $\vec{u}_t=\vec{R}_t-\vec{R}_{t-1}$, or the corresponding velocities $\vec{v}_t=\vec{u}_t/\delta t_{rec}$. These steps can be described in Cartesian or polar coordinate systems, $\vec{u}_t = (\Delta x_t , \Delta y_t) = (L_t\cos{\varphi_t},L_t\sin{\varphi_t}) = L_t  \vec{e}_t$, where $\vec{e}_t$ is a unit direction vector. The angle $\phi_t = \varphi_t-\varphi_{t-1}$ between two successive step vectors is called the turning angle. In an isotropic system, the absolute step directions, measured by the angles $\varphi_t$, cannot be of importance as well. The only relevant information of a trajectory is therefore contained in the set $\left\{ (L_t,\phi_t):t=1\ldots N\right\}$ of subsequent step lengths and turning angles. In the most general case, the underlying discrete-time random walk model has to determine the combined probability density $p\left( L_1,\phi_1,\;L_2,\phi_2,\;\ldots,\;L_N,\phi_N\right)$. In a stationary random process, the stochastic properties can only depend on the differences between the time indices. A stationary walk is therefore described by the combined probability density $p\left( \ldots,L_{t-1},\phi_{t-1},\;L_t,\phi_t,L_{t+1},\phi_{t+1},\ldots\right)$.

\section*{Aggregated statistical properties}

The aggregated statistical properties of the system are extracted by computing suitable averages. Because of the stationarity and ergodicity of the random process, we can replace ensemble averages (over different trajectories) by time averages. In the following we denote the average of a quantity $f_t$ over all absolute time points as $\left\langle f_t \right\rangle_t = \frac{1}{t_{max}\!-t_{min}\!+\!1}\sum_{t\!=\!t_{min}}^{t_{max}} f_t$.

It is important to keep in mind that, in general, a finite trajectory does not show all the symmetries of the underlying random process. For example, when analyzing relatively short trajectories with directional persistence, it may happen that all step directions fall into a narrow range of absolute angles $\varphi_t$. This can cause artifacts, such as significantly different distributions $p(\Delta x)$ and $p(\Delta y)$ of the Cartesian step components. To avoid such problems, one strategy is to use only quantities that are, by definition, invariant with respect to translations and rotations of the trajectories, such as the step length $L_t$ and the turning angles $\phi_t$.

\section*{Distributions and correlation properties of step length and turning angles}

The probability distributions of the step lengths and turning angles can be expressed formally as

\begin{eqnarray}
\label{eq:1}
p(L)&=&\left\langle \delta(L - L_t) \right\rangle_t  \nonumber\\ 
p(\phi)&=&\left\langle \delta(\phi - \phi_t) \right\rangle_t.
\end{eqnarray}

From them follow the mean value (denoted by $\overline{L}$ and $\overline{\phi}$ ) and the variances (denoted by $\sigma_L^2$ and $\sigma_{\phi}^2$). In this paper, we assume that the $\overline{L}$ and $\sigma_L^2$ are finite, excluding random walks with a heavy-tailed step length distribution such as the Levy-flight.

Besides these distributions, one should take into account possible temporal correlations of these quantities as well. The normalized autocorrelation functions of $L$ and $\phi$ (and the cross correlation function, respectively) are defined as

\begin{eqnarray}
\label{eq:2}
C_{LL}(\tau)&=&\left\langle (L_t-\overline{L})(L_{t+\tau}-\overline{L}) \right\rangle_t / \sigma_L^2 \nonumber\\
C_{\phi\phi}(\tau)&=&\left\langle (\phi_t-\overline{\phi})(\phi_{t+\tau}-\overline{\phi}) \right\rangle_t / \sigma_\phi^2 \nonumber\\
C_{L\phi}(\tau)&=&\left\langle (L_t-\overline{L})(\phi_{t+\tau}-\overline{\phi}) \right\rangle_t / (\sigma_L \sigma_{\phi}).
\end{eqnarray}

We note that in most "standard models" of random walks, such as the discrete-time correlated random walk, $L$ and $\phi$ are drawn independently from fixed distributions $p(L)$ and $p(\phi)$, so that $C_{LL}(\tau)=C_{\phi\phi}(\tau)=\delta_{\tau,0}$ and $C_{L\phi}(\tau)=0$. However, some more complex stochastic system show "super-statistical" effects, such as temporally correlated fluctuations of the step length, $C_{LL}(\tau\ge0)\neq 0$, or a time-dependent variance $\sigma_{\phi}^2=f(t)$ of the turning angle.

\section*{Vectorial velocity autocorrelation function and Mean Squared Displacement}

In contrast to the step vectors $\vec{u}_t$ themselves, certain combinations such as the dot product $\vec{u}_t\vec{u}_{t+\tau}$ are translational and rotational invariant. Therefore, a frequently used measure for the temporal structure of a random walk is the vectorial velocity autocorrelation (VAC) function:

\begin{equation}
\label{eq:3}
C_{\vec{u}\vec{u}}(\tau)=\left\langle 
(\vec{u}_t-\overline{\vec{u}})(\vec{u}_{t+\tau}-\overline{\vec{u}})
\right\rangle_t / \sigma_{\vec{u}}^2.
\end{equation}

Another popular quantity with translational and rotational invariance is the Mean Squared Displacement (MSD):

\begin{equation}
\label{eq:4}
\overline{R^2}(\tau)= \left\langle \left| 
\vec{R}_{t+\tau}-\vec{R}_t 
\right|^2 \right\rangle_t.
\end{equation}

The MSD is mathematically related to the VAC by

\begin{equation}
\label{eq:5}
\overline{R^2}(\tau)= \sigma_{\vec{u}}^2
\sum_{t=-\tau}^{t=+\tau} (\tau-t) C_{\vec{u}\vec{u}}(t) .
\end{equation}

From a practical point of view, the MSD has the advantage to be less sensitive to statistical noise, due to the summation. Also, while the normalized VAC always starts with the value 1 at lagtime zero, the MSD shows explicitly the scale of the displacements. Note that the scale factor $\sigma_{\vec{u}}^2 = \left\langle L^2 \right\rangle_t - \overline{L}^2 $ only depends on the first and second moment of the step length distribution. The shape of the MSD, on the other hand, is entirely determined by the VAC.

It is worthwhile to consider which properties of the trajectory are responsible for this shape: For a trajectory without drift ($\overline{\vec{u}}=\vec{0}$), the VAC depends only on the expression $ \left\langle \vec{u}_t\vec{u}_{t+\tau} \right\rangle_t = \left\langle L_t L_{t+\tau} \cos(\varphi_{t+\tau}-\varphi_t) \right\rangle_t $. Consider first the case when step lengths and step directions are statistically independent. Then $ \left\langle \vec{u}_t\vec{u}_{t+\tau} \right\rangle_t $ factorizes as $\left\langle L_t L_{t+\tau} \cos(\varphi_{t+\tau}-\varphi_t) \right\rangle_t \!=\! \left\langle L_t L_{t+\tau} \right\rangle_t \cdot \left\langle \cos(\varphi_{t+\tau}-\varphi_t) \right\rangle_t$. 
The first factor describes possible correlations between successive step lengths. However, the expression $\left\langle L_t L_{t+\tau} \right\rangle_t \!=\! \overline{L}^2 + \left\langle  \Delta L_t \Delta L_{t+\tau} \right\rangle_t$ is always non-equal zero, even if the step lengths fluctuations are mutually uncorrelated. The second factor describes directional correlations and it {\em can} be zero. If it {\em is} zero, the VAC is $\delta$-correlated and the MSD increases linearly with lagtime, indicating trivial diffusive behavior Consequently, any non-trivial lagtime-dependence of the VAC/MSD (such as two distinct lagtime regimes, sub-diffusive, or super-diffusive behavior) can only arise if directional correlations are present. In this case, correlated step length correlations may have an additional effect on the shape of the VAC/MSD. The most general case would even include cross-correlations between step lengths and step directions.

\section*{The projected trajectory and rotational averaging}

Next we consider aggregated statistical properties based on quantities without built-in rotational invariance. In particular, we analyze the projection of the 2D trajectory onto some axis, for example the x-axis of the coordinate system. By the projection, the sequence of vectorial steps $\vec{u}_t$ is reduced to a sequence of scalar steps $\Delta x_t$, so that some directional information is lost. However, we can define for the 1D trajectory are pair of quantities equivalent to the step length and the step direction, by factorizing each scalar step into a magnitude and a sign factor:

\begin{eqnarray}
\label{eq:6}
\Delta x_t &=& m_t \cdot s_t \nonumber\\
&=&|\Delta x_t| \cdot sgn(\Delta x_t)\nonumber\\
&=& L_t|\cos(\varphi_t)|\cdot sgn(\cos(\varphi_t)).
\end{eqnarray}

We can compute the distributions $p(m)$ and $p(s)$ of the magnitudes and signs by temporal averaging over the projected trajectory. However, in order to avoid the above-mentioned artifacts related to the finite number of steps, we have to additionally perform a rotational averaging over the absolute direction angles $\varphi$ of each step:

\begin{equation}
\label{eq:7}
\left\langle  f(\varphi) \right\rangle_{\varphi} = 
\int_0^{2\pi} \frac{d\varphi}{2\pi} f(\varphi).
\end{equation}

Using this notation, we can write

\begin{eqnarray}
\label{eq:8}
p(m)&=&\left\langle \left\langle \delta(m - m_t) \right\rangle_t  \right\rangle_{\varphi} \nonumber\\ 
p(s)&=&\left\langle \left\langle \delta(s - s_t) \right\rangle_t \right\rangle_{\varphi} .
\end{eqnarray}

We next consider how the distribution $p(m)$ is related to $p(L)$. If a single vectorial trajectory step $\vec{u}$ of step length $L$ is isotropically rotated, it produces a whole distribution $\rho(m,L)$ of projected magnitudes, ranging from $m=0$ to $m=L$:

\begin{equation}
\label{eq:9}
\rho(m,L) = \frac{ 2\;\theta(m)\;\theta(L-m)}{\pi \sqrt{L^2 \!  -\! m^2}},
\end{equation}

where $\theta()$ indicates the Heaviside step function. Therefore, a given step length distribution $p(L)$ produces a corresponding distribution of magnitudes that is given by

\begin{equation}
\label{eq:10}
p(m) = \int_m^{\infty} \!dL \; \rho(m,L) \; p(L).
\end{equation}

The quantity in 1D that corresponds to the turning angle distribution $p(\phi)$ in 2D is the probability $q=Prob("s_{t+1}\!=\!s_t")$ that two successive scalar steps have the same signs. The probability $q$ can also be called the persistence parameter of a trajectory, since $q\!=\!1/2$ indicates non-persistent behavior, while $q$-values smaller (larger) than $1/2$ indicate sub-diffusive (super-diffusive) behavior In order to derive a relation between $p(\phi)$ and $q$, we consider a sequence of two vectorial steps $\vec{u}_1$ and $\vec{u}_2$, enclosing a turning angle $\phi$ (compare Fig.2) with probability $p(\phi)d\phi$. Assume that initially the two sign factors $s_1$ and $s_2$ are both positive. If we now gradually increase $\phi$, there is a critical turning angle $\phi_c(\varphi)$ at which $s_1=s(\varphi)=sgn(\cos(\varphi))$ becomes different from $s_2 =s(\varphi+\phi)=sgn(\cos(\varphi+\phi))$. We can therefore express $q$ as an angular integral over $p(\phi)$:

\begin{eqnarray}
\label{eq:11}
q &=& \int_0^{2\pi} \frac{d\varphi}{2\pi} \;\int_{-\pi}^{+\pi} \! d\phi  \; \delta_{s_1,s_2} \; p(\phi) \;\;\mbox{with}\nonumber\\
&=& \int_{-\pi}^{+\pi} \! d\phi \; \left[ \int_0^{2\pi} \frac{d\varphi}{2\pi} \; \delta_{s(\varphi),s(\varphi+\phi)} \right] \; p(\phi)\nonumber\\
&=& \int_{-\pi}^{+\pi} \! d\phi \; \left[\;1-|\phi/\pi|\;\right] \; p(\phi).
\end{eqnarray}

The fact that $p(\phi)$ is a distribution, whereas $q$ is just a number, demonstrates the information loss associated with the projection from 2D to 1D. This missing information is the detailed shape of the turning angle distribution, which in many cases will not be of particular interest. In this sense, the 1D projection of a 2D trajectory with rotational averaging represents a useful simplification and helps to filter out the essential information.

Finally, the temporal structure of the projected trajectory can be analyzed with the rotationally averaged autocorrelation function of the scalar steps:

\begin{equation}
\label{eq:12}
C_{\Delta x\Delta x}(\tau)=\left\langle \left\langle
(\Delta x_t-\overline{\Delta x})(\Delta x_{t+\tau}-\overline{\Delta x})
\right\rangle_t / \sigma_{\Delta x}^2 \right\rangle_{\varphi}.
\end{equation}

The correlation functions $C_{mm}(\tau)$, $C_{ss}(\tau)$ and $C_{ms}(\tau)$ are defined in an analogous way.

\section*{Discrete pattern statistics and the persistent Markov chain of signs}

By the above procedure, we have mapped an originally two-dimensional trajectory onto two scalar time series, $m_t$ and $s_t$. Since the signs $s_t$ are binary variables, we can apply to them analysis tools that are tailor-made for discrete random processes. In particular, we can count the frequency of patterns, such as "-+-", within the time series. Once the probabilities for all patterns of a given length are known, it is straight forward to construct a higher order Markov model that replicates the statistical properties of a measured time series.

The principles of pattern statistics can be demonstrated with a simple binary Markov chain $s_t$: At $t=0$ it starts randomly with "-" or "+" (equal probability). For each subsequent time, $Prob("s_{t+1}\!=\!s_t")=q$, with a pre-defined persistence parameter. What is the frequency distribution for a given pattern, such as "-+-", in this model ? In this particular case, $p("-+-")=p("-")p("-"\rightarrow"+")p("+"\rightarrow"-")$, which yields $p("-+-")=(1/2)\;(1-q)\;(1-q)$. For reasons of symmetry, the probability of any pattern is equal to that of its inverse, where all "+" and "-" are exchanged. Also, we can temporally reverse a pattern without changing its probability. Thus, there are only 3 distinct patterns of length 3, and their relative frequencies in our model are: $p("---")\propto q^2$, $p("--+")\propto q(1-q)$ and $p("-+-")\propto (1-q)^2$. They all become equally frequent for the non-persistent case $q=1/2$.

\section*{The restricted turning angle (RTA) model}

We next consider the class of 2D random walk models in which the step lengths $L$ and step directions $\varphi$ are statistically independent, $\left\langle  L_t \varphi_{t+\tau} \right\rangle_t=0$. The step lengths have a fixed distribution $p(L)$ with finite values for mean $\overline{L}$ and variance $\sigma_L^2$. The turning angles also have a fixed distribution $p(\phi)$, with zero mean and variance $\sigma_{\phi}^2$. In particular, we are interested in the case where $\sigma_{\phi}^2$ is rather narrow (restricted turning angles), so that the walk has directional persistence. We call this case the RTA model in the following.

It is straight forward to show that under the given assumptions the vectorial velocity autocorrelation functions is given as

\begin{equation}
\label{eq:13}
C_{\vec{u}\vec{u}}(\tau=n \;t_{rec}) =
\left[ \frac{\overline{L}^2+\sigma_L^2\delta_{n0}}{\overline{L}^2+\sigma_L^2\;\;\;} \right]\;
\left\langle \cos(\varphi_{t+n}-\varphi_{t}) \right\rangle_t
\end{equation}

The directional correlation factor can be expressed as an integral over turning angles:

\begin{equation}
\label{eq:14}
\left\langle \cos(\varphi_{t+n}-\varphi_{t}) \right\rangle_t =
\int_{-\pi}^{+\pi} d\phi\; p(\phi,n) \cos(\phi).
\end{equation}

Here, $p(\phi,n)$ is the probability density that a vectorial step and its $n$th successor enclose a turning angle $\phi$. It is clear that $p(\phi,n=0)=\delta(\phi-0)$, that $p(\phi,n=1)=p(\phi)$ is just the prescribed turning angle distribution and that $p(\phi,n\rightarrow\infty)\rightarrow 1/(2\pi)$. The temporal development of $p(\phi,n)$ corresponds to a kind of diffusion process on the unit circle. As long as the width of $p(\phi,n)$ is smaller than $2\pi$, we can view the process as a diffusion on a linear $\phi$-axis. Then, $p(\phi,n)$ is just the $n$-fold convolution of the turning angle distribution $p(\phi)$ with itself. For lagtimes $1\ll n\ll n_{max}=2\pi/\sqrt{\sigma_{\phi}^2}$, the distributions $p(\phi,n)$ resemble normalized Gaussians with zero mean and a lagtime-dependent variance $\sigma_{\phi}^2(n)=n\cdot\sigma_{\phi}^2$.  We insert the approximation $p(\phi,n)\approx\frac{1}{\sqrt{2\pi\sigma_{\phi}^2(n)}}\;e^{-(1/2)\phi/\sigma_{\phi}^2(n)}$ into Eq.[14] and obtain analytically

\begin{equation}
\label{eq:15}
\left\langle \cos(\varphi_{t+n}-\varphi_{t}) \right\rangle_t \approx
e^{-(\sigma_{\phi}^2/2)\;n}.
\end{equation}

Summing up, the velocity autocorrelation function in the RTA model will show a sudden drop between lagtimes $0$ and $1$. This drop occurs because the variance of the uncorrelated step lengths $L_t$ contributes only to the total velocity variance ($n=0$). For intermediate lagtimes in the regime $1\ll n\ll n_{max}=2\pi/\sqrt{\sigma_{\phi}^2}$ it will decay exponentially with a characteristic decay time inversely proportional to the variance of the turning angle distribution:

\begin{equation}
\label{eq:16}
C_{\vec{u}\vec{u}}(n) \approx
\left[ \frac{\overline{L}^2+\sigma_L^2\delta_{n0}}{\overline{L}^2+\sigma_L^2\;\;\;} \right]\;
e^{-(\sigma_{\phi}^2/2)\;n}.
\end{equation}

\section*{Simulation of RTA model}

For a concrete example, we consider Rayleigh-distributed step lengths with the most probable value $L_p$, $p(L\ge 0)=(L/L_p^2)\;e^{-L^2/(2L_p^2)}$, so that $\overline{L}=\sqrt{\pi/2}L_p$ and $\sigma_L^2=\frac{4-\pi}{2}L_p^2$. The turning angles are assumed to be equally distributed within a narrow interval $\left[ -\phi_{max}\ldots +\phi_{max} \right]$, corresponding to a variance $\sigma_{\phi}^2=\frac{1}{3}\phi_{max}^2$. Note that this model has only the two parameters $L_p$ and $\phi_{max}$.

For the numerical simulation of the RTA model we set $L_p=1.0$lu, with an arbitrary length unit lu, and $\phi_{max}=\pi/20$. The recording time interval is set $\delta_{rec}=1$. A segment of a typical trajectory is shown in Fig.3. The numerically calculated step width distributions, turning angle distributions and velocity autocorrelation function, together with the analytical results, are shown in Figs.4-5.

\section*{1D-projection of RTA model}

Using the transformation formula Eq.10, we obtain for the 1D projection of the RTA model with Rayleigh-distributed step lengths a Gaussian magnitude distribution $ p(m\ge 0)=\frac{2}{L_p\sqrt{2\pi}} \; e^{-(L/L_p)^2/2}$ (see Fig.6). According to Eq.11, we expect a persistence parameter $q=1-\frac{\phi_{max}}{2\pi}$, yielding $q=0.975$ for a maximum turning angle $\phi_{max}=\pi/20$. By counting the fraction of pairs of identical signs in the simulated time series $s_t$ from the projected RTA model, we obtain $q=0.974$.

The autocorrelation function of $s_t$ is shown in Fig.7. It is interesting to compare it with $C_{ss}(n)$ for a persistent Markov chain of signs with the same average $q$, which decays like $C_{ss}(n)=(2q-1)^n$ for $q \ge 1/2$. Note that the two models agree only for $n=0$ (automatic due to normalized autocorrelation function) and $n=1$ (correct average fraction of equal sign pairs). For larger lagtime, the sign correlations in the projected RTA model decay more slowly than in the Markov chain.

The origin of these deviations are "higher order" correlations in the projected RTA model, beyond a simple Markov approximation. Consider the projection of the RTA-trajectory onto the x-axis. As long as the step direction is roughly parallel to the x-axis, the time series $s_t$ shows long sequences of identical signs, such as "+++++++", what could also be called a "bunching" of equal sign pairs. But occasionally, the direction diffuses into a close to vertical position. During such phases, $s_t$ shows sequences such as "+-+-++-+--+", corresponding to an "anti-bunching" of equal sign pairs. Thus, while the average fraction of equal sign pairs agrees in both models, they are spread over the time axis in a different way in the projected RTA model. These differences are also reflected in the pattern statistics: For example, in the projected RTA model (Fig.9), the fraction of "--+" patterns is significantly diminished, compared to the Markov chain (Fig.8).

\section*{A "momentary persistence" variable and its temporal correlations}

The temporal distribution of equal sign pairs can be investigated by defining a momentary persistence variable,

\begin{equation}
\label{eq:17}
\eta_t = \delta_{s_{t\!-\!1},s_t}.
\end{equation}

The global persistence parameter $q=\left\langle \eta_t \right\rangle_t$ is just the time average of this variable. In a persistent Markov chain of signs, the random variable $\eta_t$ behaves like a Bernoulli process with the probability $q$ for the event $1$ and $1-q$ for the event $0$. For such a white noise process, the autocorrelation function yields $C_{\eta\eta}(\tau)=\delta_{\tau,0}$. In the projected RTA model, however, the momentary persistence should have some memory time larger than zero. This is indeed the case, as shown in Fig.10.

\begin{figure}[htb] \includegraphics[width=14cm]{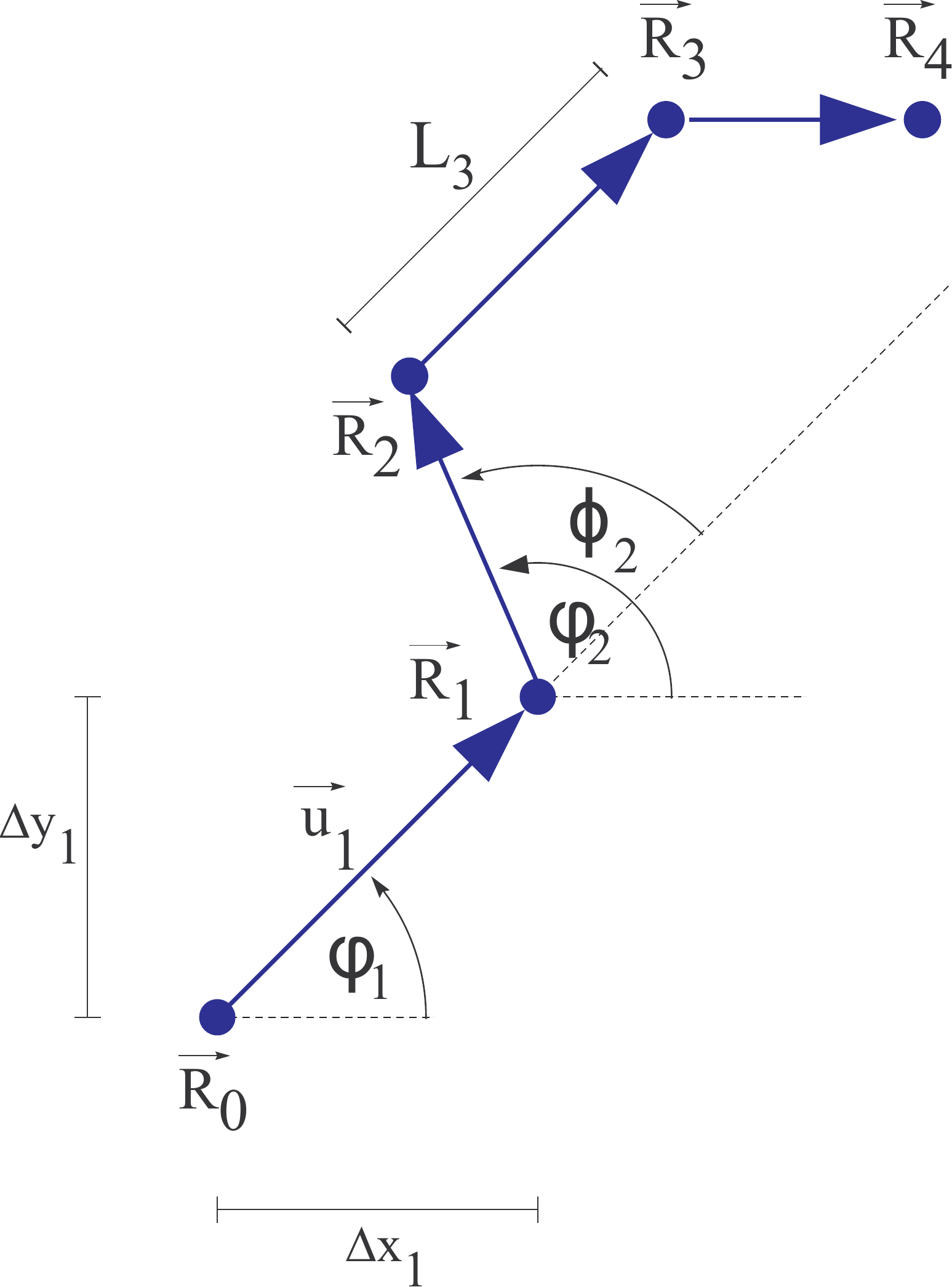} \caption{\label{1} {\em 
Sample trajectory, introducing positions $\vec{R}_t$, steps $\vec{u}_t$, step directions $\varphi_t$, step lengths $L_t$ and turning angles $\phi_t$.
}}\end{figure}

\begin{figure}[htb] \includegraphics[width=14cm]{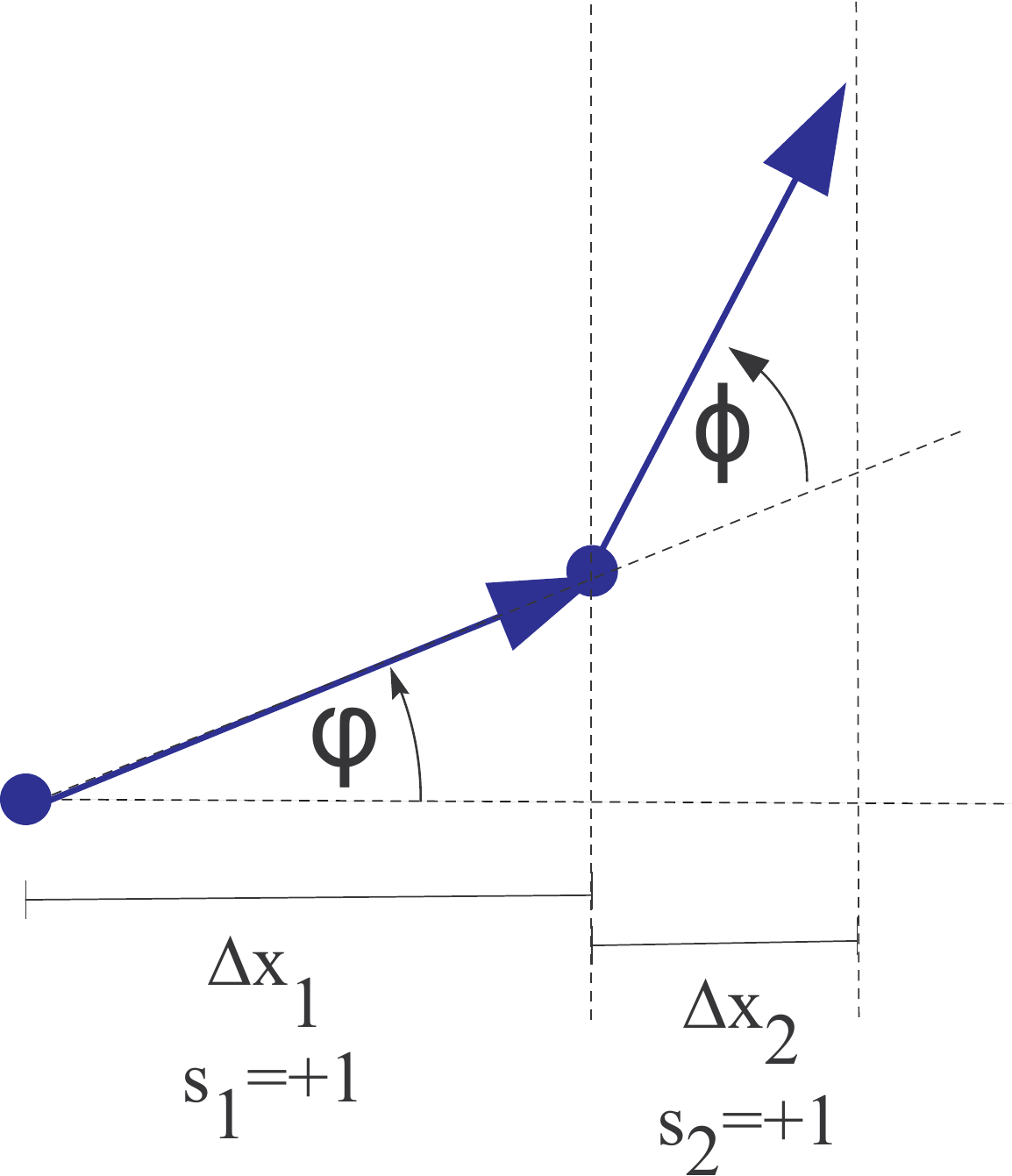} \caption{\label{2} {\em 
Two successive vectorial steps, their projections $\Delta x_t$ onto the x-axis, and the signs $s_t$ of the steps.
}}\end{figure}

\begin{figure}[htb] \includegraphics[width=14cm]{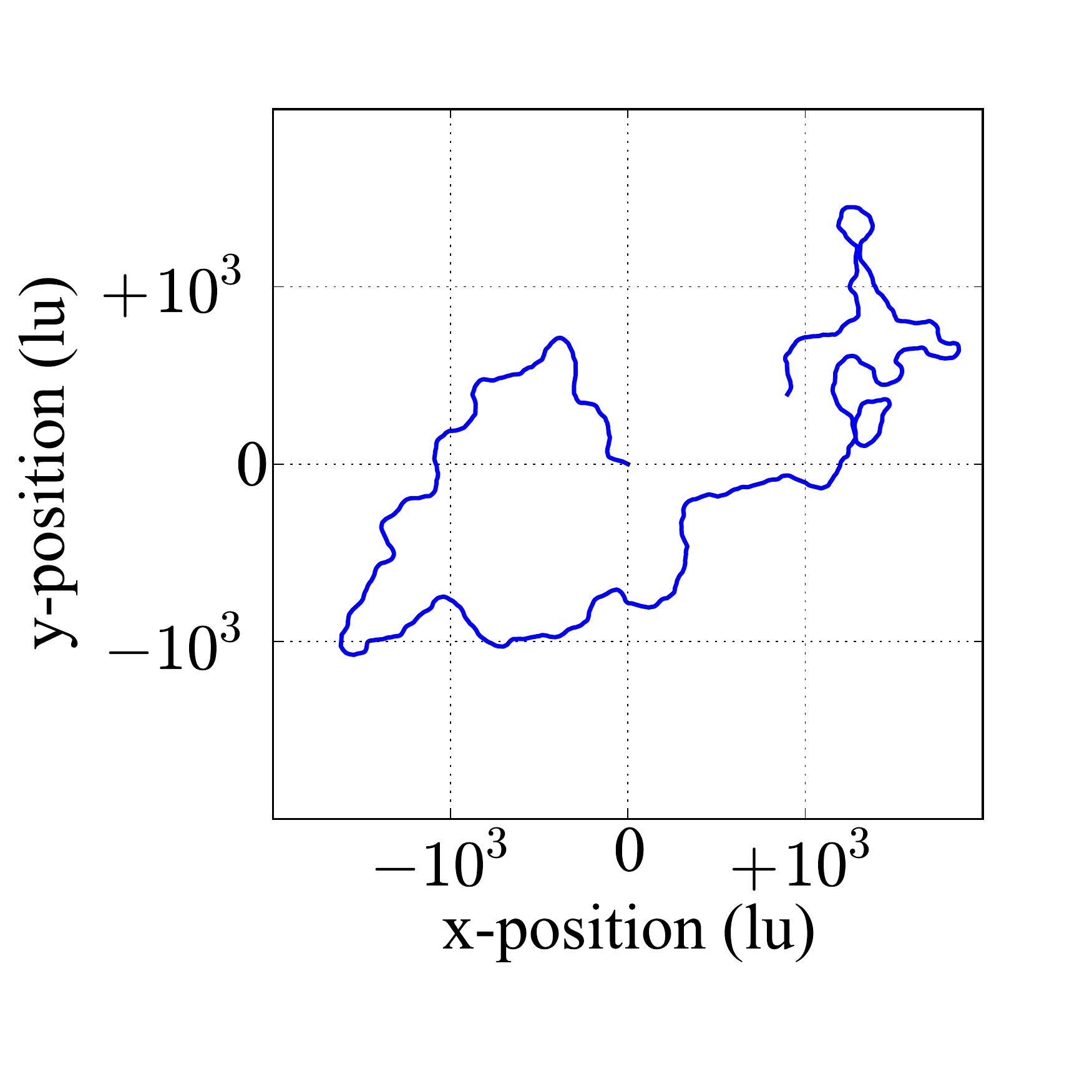} \caption{\label{3} {\em 
Segment of a trajectory in the RTA model.
}}\end{figure}

\begin{figure}[htb] \includegraphics[width=14cm]{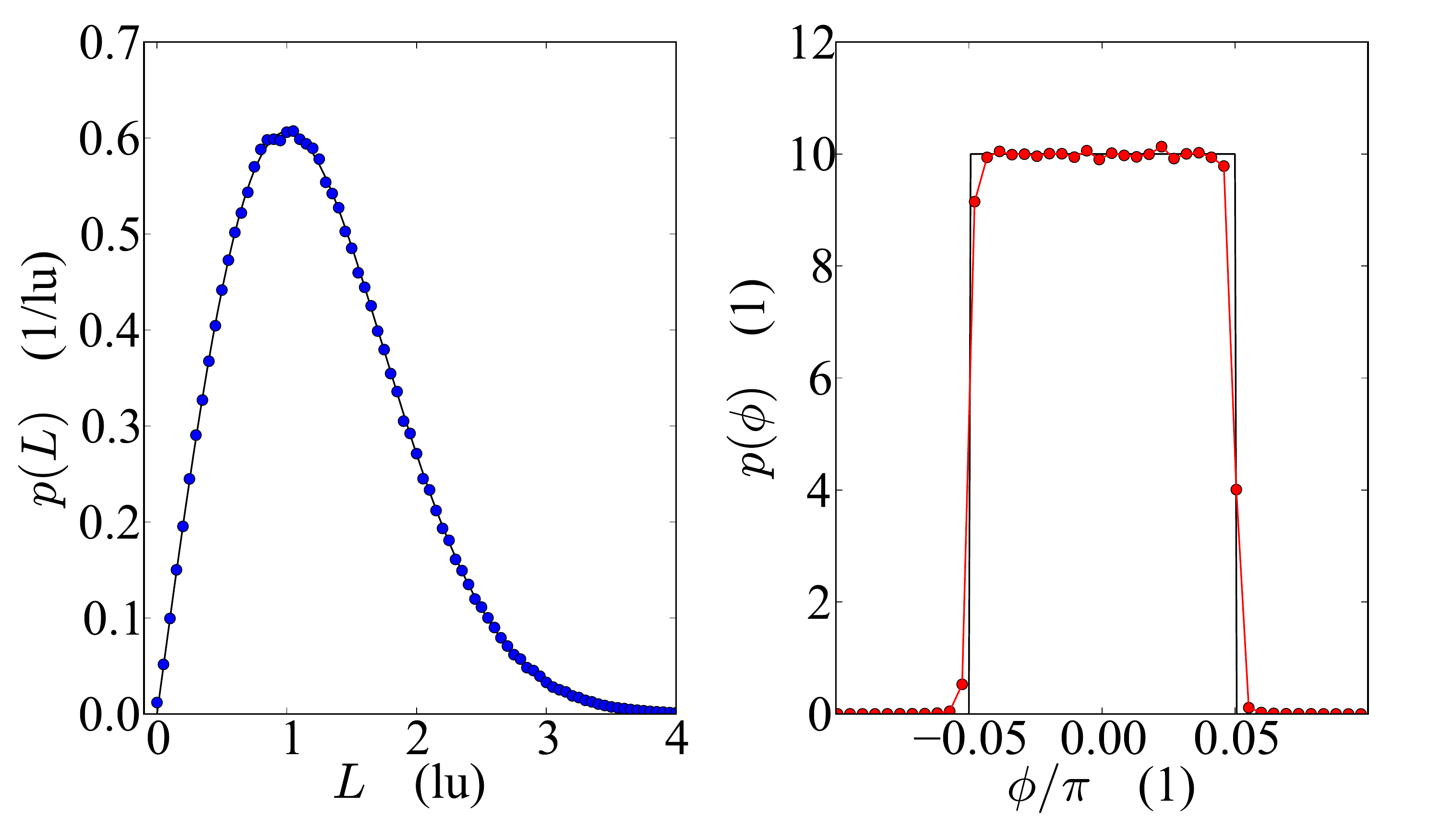} \caption{\label{4} {\em 
Simulated (symbols) and analytic (black lines) distributions of step lengths (left) and turning angles (right) in the RTA-model.
}}\end{figure}

\begin{figure}[htb] \includegraphics[width=14cm]{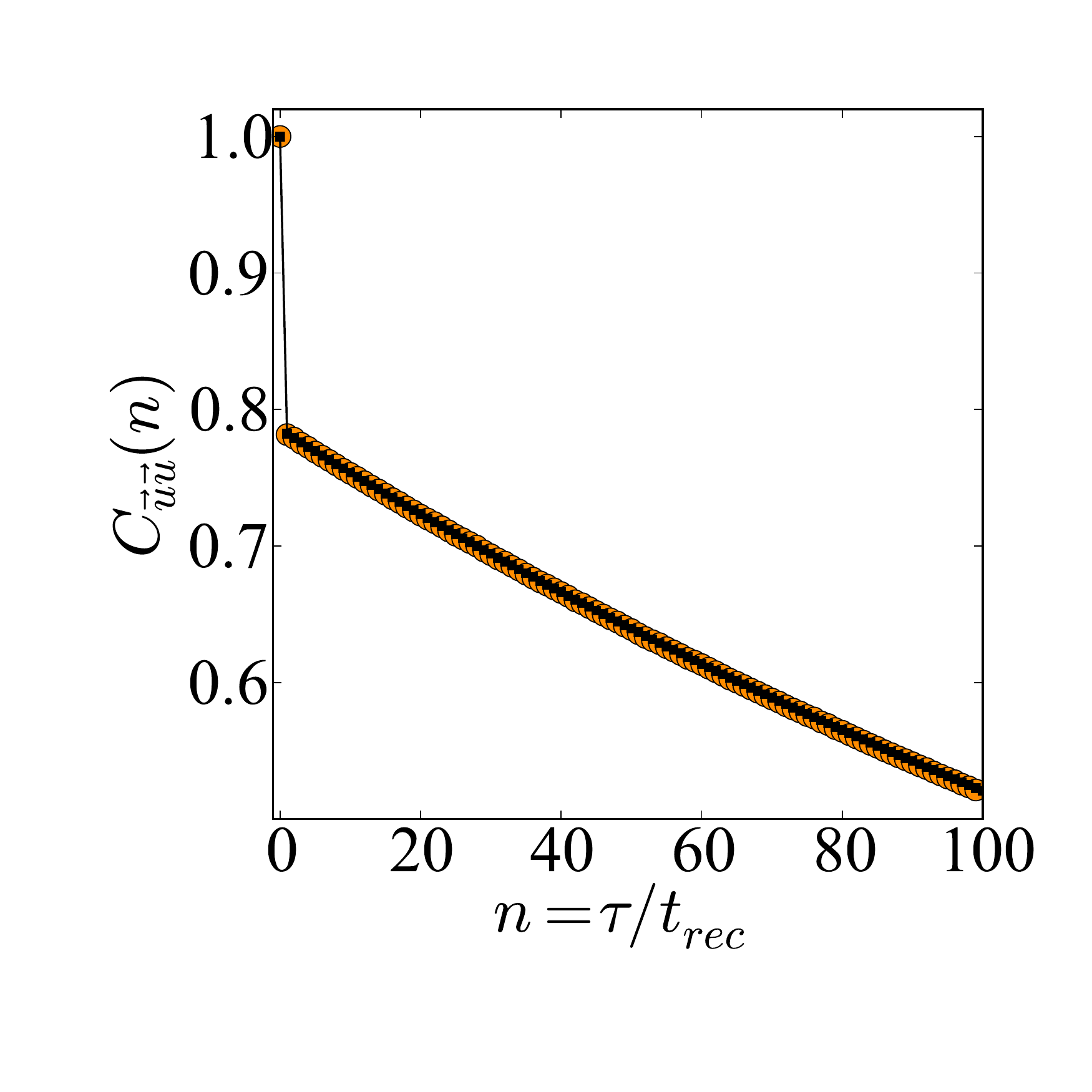} \caption{\label{5} {\em 
Vectorial velocity autocorrelation function in the RTA-model, comparing simulation (red symbols) with analytical approximation (black line).
}}\end{figure}

\begin{figure}[htb] \includegraphics[width=14cm]{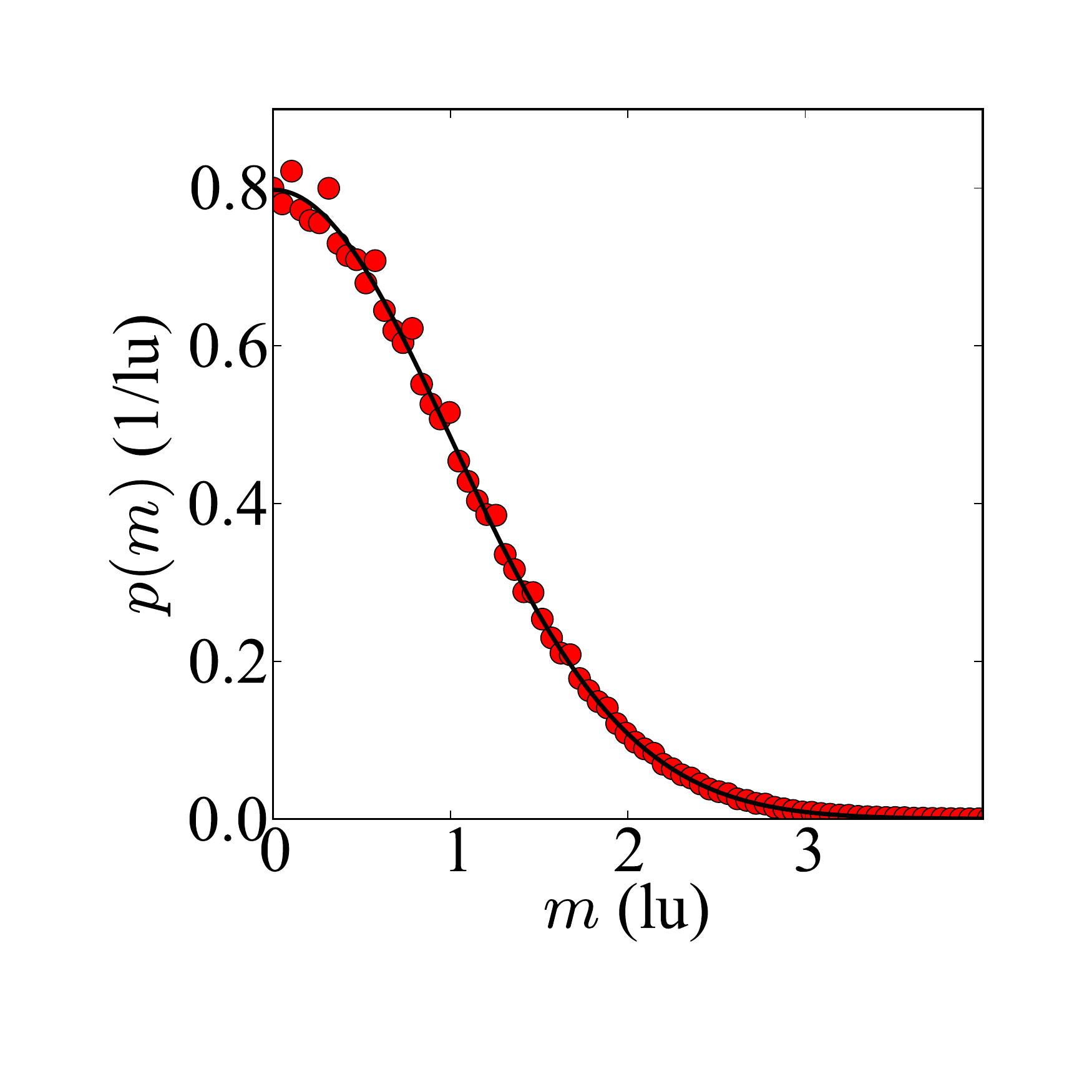} \caption{\label{6} {\em 
Simulated (symbols) and analytic (black line) distribution of step magnitude in the projected RTA-model.
}}\end{figure}

\begin{figure}[htb] \includegraphics[width=14cm]{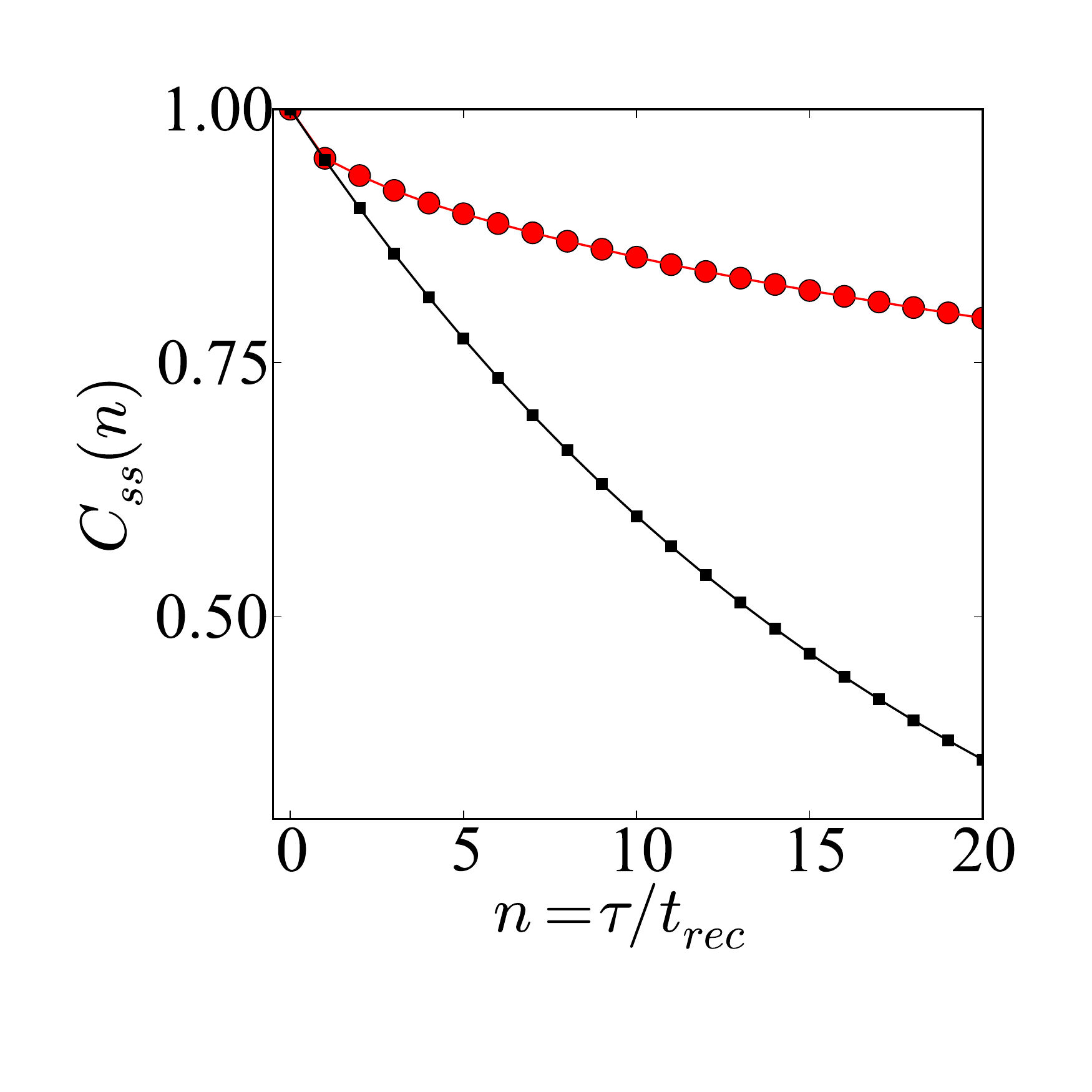} \caption{\label{7} {\em 
Autocorrelation functions of sign factors in the projected RTA-model (red), and in a persistent Markov chain of signs (black). Note that models agree only at lagtimes $0$ and $1$.
}}\end{figure}

\begin{figure}[htb] \includegraphics[width=14cm]{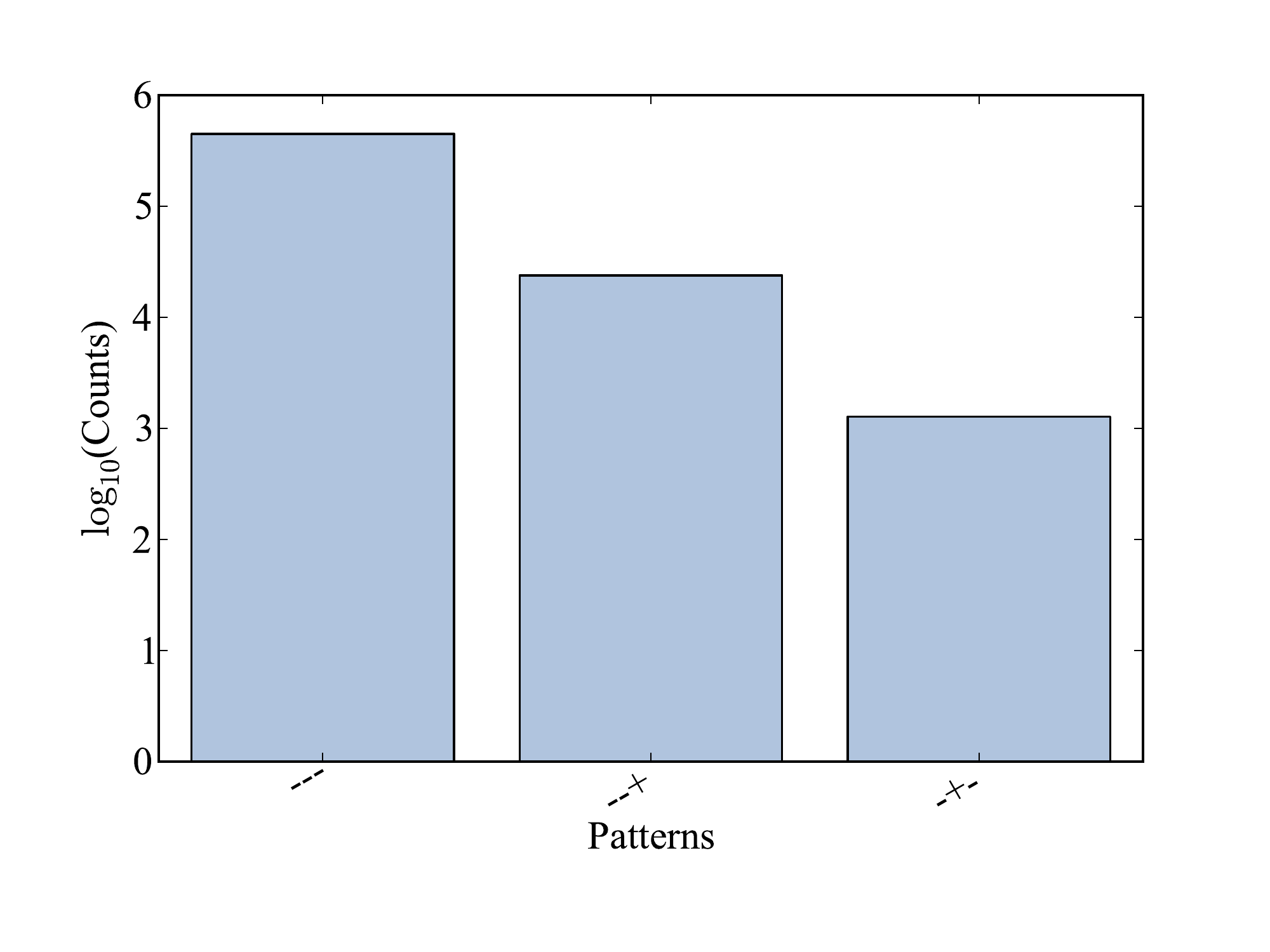} \caption{\label{8} {\em 
Logarithmic frequency of selected patterns in a persistent Markov chain of signs $s_t$, for $q=0.975$.
}}\end{figure}

\begin{figure}[htb] \includegraphics[width=14cm]{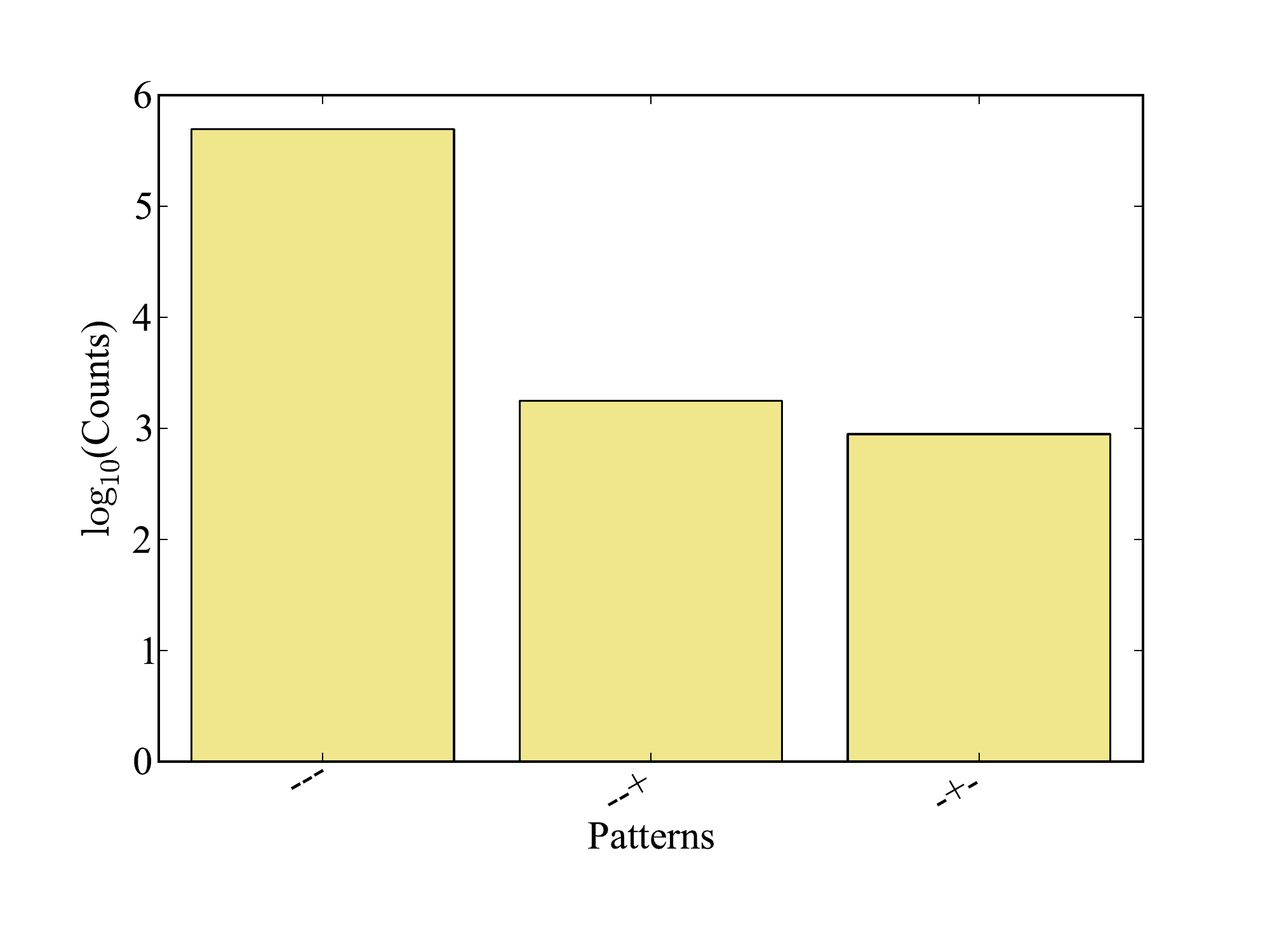} \caption{\label{9} {\em 
Logarithmic frequency of selected patterns in the projected RTA model.
}}\end{figure}

\begin{figure}[htb] \includegraphics[width=14cm]{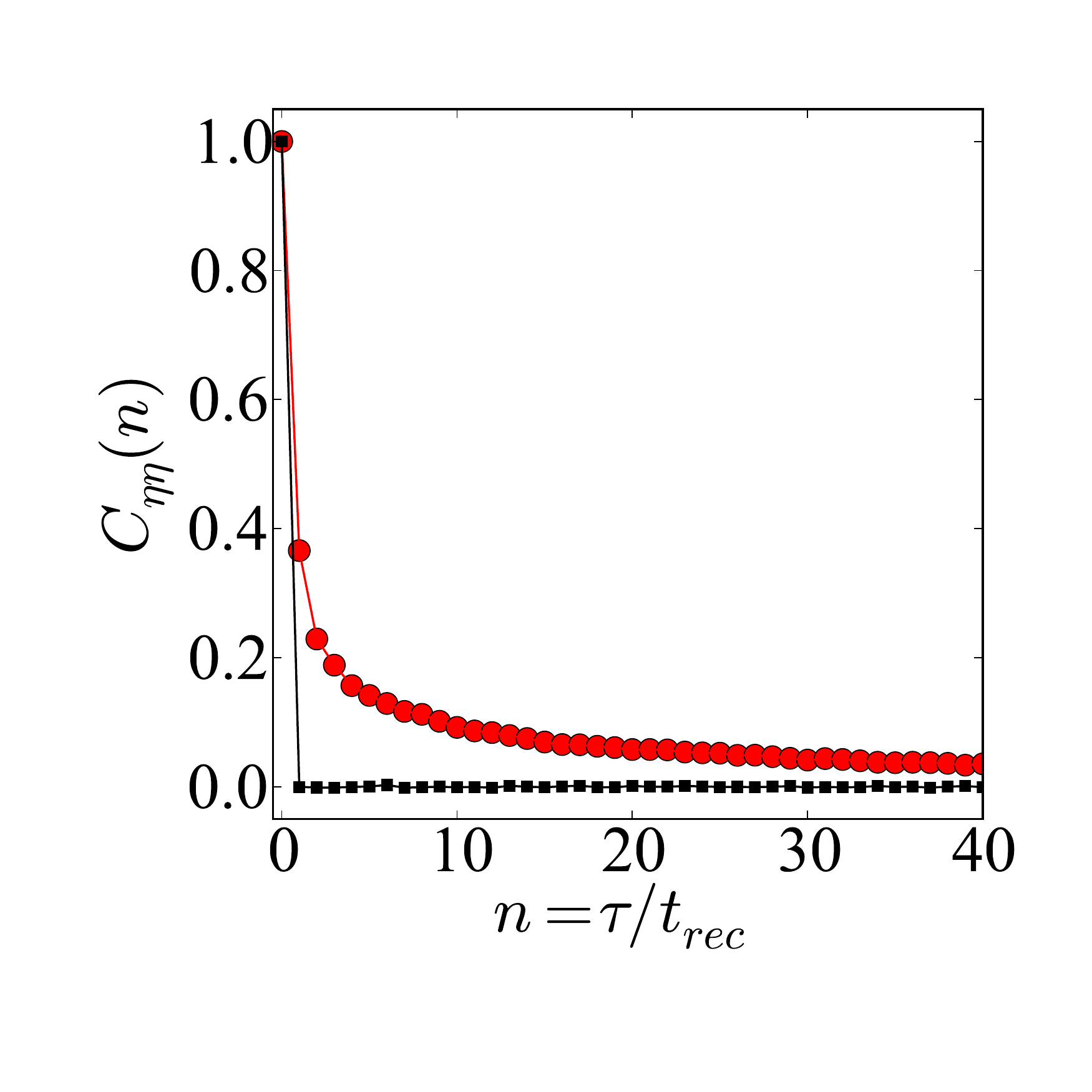} \caption{\label{10} {\em 
Autocorrelation function of the momentary persistence $\eta_t$ in the projected RTA model (red) and in a persistent Markov chain of signs with the same $q$ (black).
}}\end{figure}

\begin{acknowledgments}
This work was supported by grants from Deutsche Forschungsgemeinschaft.
\end{acknowledgments}
\bibliography{refs}
\end{document}